\documentclass[USenglish, 12pt]{article}	
\usepackage[utf8]{inputenc}				%(only for the pdftex engine)
\usepackage[big,online]{dgruyter}	%values: small,big | online,print,work
\usepackage{lmodern} 
\usepackage{microtype}
\usepackage[round]{natbib}
\usepackage{setspace}

\theoremstyle{dgthm}

\theoremstyle{dgdef}

\begin{document}

%%%--------------------------------------------%%%
	\articletype{Research Article}
	\received{Month	DD, YYYY}
	\revised{Month	DD, YYYY}
  \accepted{Month	DD, YYYY}
  \journalname{}
  \journalyear{YYYY}
  \journalvolume{XX}
  \journalissue{X}
  \startpage{1}
  \aop
  \DOI{10.1515/sample-YYYY-XXXX}
%%%--------------------------------------------%%%

\title{Predicting Batting Averages in Specific Matchups Using Generalized Linked Matrix Factorization}
\runningtitle{Predicting Batting Averages Using GLMF}
%\subtitle{Insert subtitle if needed}

\author*[1]{Michael J. O'Connell}
%\ use * to mark the author as the corresponding author
%\author[2]{Second Author}
%\author[2]{Third Author} 
\runningauthor{O'Connell}
\affil[1]{\protect\raggedright 
}
%\affil[2]{\protect\raggedright 
%Institution, Department, City, Country of second author and third author, e-mail: author\_two@xx.yz, author\_three@xx.yz}
	
%\communicated{...}
%\dedication{...}
	
\abstract{Predicting batting averages for specific batters against specific pitchers is a challenging problem in baseball. Previous methods for estimating batting averages in these matchups have used regression models that can incorporate the pitcher's and batter's individual batting averages. However, these methods are limited in their flexibility to include many additional parameters because of the challenges of high-dimensional data in regression. Dimension reduction methods can be used to incorporate many predictors into the model by finding a lower rank set of patterns among them, providing added flexibility. This paper illustrates that dimension reduction methods can be useful for predicting batting averages. To incorporate binomial data (batting averages) as well as additional data about each batter and pitcher, this paper proposes a novel dimension reduction method that uses alternating generalized linear models to estimate shared patterns across three data sources related to batting averages. While the novel method slightly outperforms existing methods for imputing batting averages based on simulations and a cross-validation study, the biggest advantage is that it can easily incorporate other sources of data. As data-collection technology continues to improve, more variables will be available, and this method will be more accurate with more informative data in the future.}

\keywords{Data Integration, Dimension Reduction, Sabremetrics, Missing Data Imputation}

\maketitle

\doublespacing

\section{Introduction} 

Sabremetrics, the quantitative analysis of baseball data, has been gaining interest due to improved technology for data collection and growing interest in analytics \citep{lewis2004moneyball}. One challenge in sabremetrics is predicting the performance of specific batters against specific pitchers. For some matchups, there may be sufficient past data to obtain a reasonable estimate of the expected batting average (for instance, veteran players who are regular starters and play in the same division). But for most matchups, the sample size of previous matchups is very small or non-existent. When the sample size is very small, the estimate is unreliable because it is subject to very high variability. When a batter has never faced a pitcher before, there is no empirical data to directly estimate the expected batting average for that matchup. With the number of players in the league, this is very common. For instance, in the 2017 Major League Baseball (MLB) data that was used for this analysis, 76\% of the possible matchups never occurred during that season. A more complex method that leverages other information about the players involved is necessary to predict the batting averages for these pairings. 

One approach that has been used to predict batting averages is the log5 model \citep{james1983abs}. This idea behind this model is very simple: it takes into account the batter's overall average, the pitcher's overall average, and the overall league average. Other papers have used this method in predictive modeling. \cite{healey2015modeling} adapted this model to predict the probability of a strikeout in a specific batter/pitcher matchup. \cite{doo2018modeling} used a Bayesian approach to improve the predictions of the model. Both of these papers incorporated other covariates into the model, such as the groundouts. But many potential covariates are available for each batter and pitcher, and adding that information into these models requires increasing the number of parameters in the model. This can be a challenge when sample sizes are small or when the number of covariates to include in the model is large. 

This paper introduces a multi-source dimension reduction approach to estimating batting averages in specific matchups. Multi-source data are high-dimensional data sets that consist of multiple different data types that have at least one shared dimension, which can be either a shared variable set or a shared sample set. There are two types of multi-source data, depending on the shared dimension. Vertical integration refers to data sets that share their column labels. The shared columns typically represent multiple different data types for a shared set of subjects. In horizontal integration, the data sets share their row labels. The shared rows usually represent a shared set of variables over different sets of subjects. In this analysis there is bidimensional integration, which means that both horizontal and vertical integration are present. The batter versus pitcher matchup data is set up such that the columns represent different pitchers and the rows represent different batters. There is also aggregate batting data for the batters and aggregate pitching data for the pitchers. The aggregate batting data is horizontally integrated with the matchup data, while the aggregate pitching data is vertically integrated with the matchup data.

Multi-source data have become much more common as technology has allowed for the collection and storage of several types of high-throughput data. The biggest challenge in multi-source data is that the data sources are related, so analyzing them separately can miss important associations between them. Many methods have been developed to identify features that are shared across different data sources \citep{lock2013joint, lofstedt2011onpls, yang2016non, zhou2015group}. However, these methods are usually limited to data sets that only contain vertical integration or horizontal integration. A few recent methods have been developed to account for bidimensionally linked data sets. \cite{o2019linked} introduced a method for data sets where one matrix shares its column labels with one data set and its row labels with another data set. \cite{park2020integrative} extended this concept to data sets with four matrices that each share row labels with one matrix and column labels with another and allowed more complex shared structure. \cite{yuan2021double} introduced a similar method for double-matched matrices, which share both their row and column labels with each other.   

The linked matrix factorization (LMF) method described in \cite{o2019linked} best matches the data structure in this analysis. LMF is appropriate for data sets where one matrix (X) is vertically integrated with another matrix (Y) and horizontally integrated with a different matrix (Z). An illustration of this data structure in the context of the baseball data can be seen in Figure 1. The LMF algorithm uses an alternating least squares approach to estimate the scores and loadings. However, this approach requires that the data are quantitative and works best when the data are normally distributed because it is based on ordinary least squares estimates for the scores and loadings. Thanks to the central limit theorem, it is often reasonable to assume that the mean follows a normal distribution. However, some contexts may be much better served with more accurate distributional assumptions. For example, batting averages in specific batter versus pitcher matchups are binomial observations, and most of the observations have a very small sample size. It is not reasonable to invoke the normal approximation of the binomial distribution in this case because the sample sizes are small and the distribution is right-skewed. 

Generalized Linear Models (GLMs) are one way to adapt least squares estimation to non-Gaussian data. GLMs are used when data are assumed to follow some exponential family distribution.  Exponential familiy distributions are distributions that can be parameterized as follows:
$$ f(x \vert \theta) = h(x)\exp{\{x\theta - b(\theta)\}}. $$
In this parameterization, $\theta$ is known as the natural parameter and is based on a function of the mean, $h(x)$ is some function of $x$ that does not depend on $\theta$, and $b(\theta)$ is some function of $\theta$ that does not depend on $x$. 
Common examples of exponential family distributions are the normal, binomial, and Poisson distributions. 
A key component of the GLM model is the link function, which is a function that relates the natural parameter $\theta$ and the mean $\mu = E[x]$. Under the exponential family model, $\mu = b'(\theta)$ and $\text{Var}[x] = b''(\theta)$. The canonical link function is defined as $g(\mu) = b'^{-1}(\theta)$. Commonly used canonical link functions are shown in Table \ref{tab:expfam}.

 \begin{table} [!ht]
  %\centering
    \caption[Common exponential family distributions and their corresponding canonical links]{Common exponential family distributions and their corresponding canonical links.}
  \begin{tabular}{ c c c c }
    %\hline
    Distribution & f(x) & $\mu$ & Canonical Link \\
    \hline
    Normal & $ (2\pi\sigma^2)^{-\frac{1}{2}}e^{(x-\mu)^2} $ & $\theta$ & Identity ($g(\mu)=\mu$) \\
    Binomial & $ {N\choose x}\theta^x(1-\theta)^{N-x} $ & $\frac{e^\theta}{1+e^\theta}$ & Logit ($g(p) = \log\frac{p}{1-p}$) \\
    Poisson & $ \frac{1}{\theta}e^{-\theta x} $ & $e^\theta$ & Log ($g(\lambda) = \log{\lambda}$) \\
    %\hline
  \end{tabular}
  \label{tab:expfam}
\end{table}

GLMs are fit using Iteratively Reweighted Least Squares (IRLS) \citep{green1984iteratively}. IRLS is a method that maximizes the likelihood of a GLM by alternating between fitting a weighted least squares regression and adjusting the weights. Some methods have been developed to handle matrix decompositions of non-Gaussian data. An exponential family version of PCA was developed to maximize the likelihood under exponential family models \citep{collins2002generalization}. In the exponential PCA model, it is assumed that each entry $x_{ij}$ of a matrix $X$ follows a particular exponential family distribution given $\Theta_{ij}$. Exponential PCA decomposes the natural parameter matrix $\Theta$ into loadings $U$ and scores $V$:
$$ \Theta = UV^T $$

In the multi-source context, \citet{li2018general} introduced the Generalized Association Study (GAS), which is used for heterogeneous multi-source data with one shared dimension. Heterogenous data are data that follow different distributional assumptions. GAS allows the different data sources to follow different distributional assumptions. Their method models a decomposition of the natural parameter space rather than a decomposition of the mean. GAS is also uses an alternating least squares approach, using an IRLS algorithm that can accommodate heterogeneous data to estimate either the scores and loadings for the decomposition at each step. 

The motivating example of this project is a batter vs. pitcher data set, which had the number of at-bats (AB) and hits (H) between each specific batter and pitcher combination in Major League Baseball (MLB). The model assumes that these data can be represented with a binomial distribution: $ H \sim Bin(AB, p) $, where p represents the hypothetical true batting average for a particular batter against a particular pitcher.  The analysis incorporates the pitching stats for all of the pitchers, as well as the batting stats for all of the batters. As mentioned previously, this is the same data structure used in the LMF algorithm; in this case the batter vs. pitcher data is X, the pitching data is Y, and the batting data is Z (Figure 1). The complication is that the data are heterogenously distributed. Because most batters had few to no at-bats against specific pitchers, it makes more sense to assume that X follows a binomial distribution. Although many of the statistics incorporated in Y and Z are also proportions (after accounting for the number of plate appearances or innings pitched), the number of trials for these observations are much larger, so it is reasonable to assume that these are normally distributed. 

This paper introduces a linked matrix factorization method for generalized linear models. The resulting algorithm, Generalized Linked Matrix Factorization (GLMF), allows for the simultaneous dimension reduction of bidimensionally integrated data sets that follow potentially heterogeneous exponential family distributions. The performance of this method is then compared to other methods for predicting batting averages of specific batter versus pitcher matchups. 

\section{Integrative Factorization} \label{mod1}

The three matrices involved in the GLMF are $X$, $Y$, and $Z$, where $X$ shares its row space with $Y$ and its column space with $Z$ (Figure 1). Assume $X$, $Y$, and $Z$ each follow some exponential family distribution, although each of them can follow different distributions. Let the dimensions of X be $m_1\times n_1$, the dimensions of Y be $m_2\times n_1$, and the dimensions of Z be $m_1 \times n_2$.  The goal of the decomposition is to leverage shared structure across $X$, $Y$, and $Z$ in a simultaneous low-rank factorization. 

Begin by assuming exponential family distributions ($F_x$, $F_y$, and $F_z$) on each of the data sets.  
\begin{align*}
X \sim F_X(\Theta_X) \\ 
Y \sim F_Y(\Theta_Y) \\
Z \sim F_Z(\Theta_Z)
\end{align*}

In the context of the baseball data, assume
\begin{align*}
X \sim Bin(N, P) \\ 
Y \sim N(\mu_Y, \sigma^2_Y) \\
Z \sim N(\mu_Z, \sigma^2_Z)
\end{align*}
where $ P = Inverse Logit(\Theta_X) $, $ \mu_Y = \Theta_Y $, and $ \mu_Z = \Theta_Z$. 

Define a joint rank $r$ approximation for the three data matrices as follows: 
\begin{align*}
\Theta_X &= UV^T \\ 
\Theta_Y &= U_yV^T \\
\Theta_Z &= UV_z^T 
\end{align*}
where each $\Theta$ represents the corresponding natural parameter matrix.
\begin{itemize}
\item $U$ is an $m_1 \times r$ matrix representing the row structure shared between $\Theta_X$ and $\Theta_Z$
\item $V$ is an $n_1 \times r$ matrix representing the column structure shared between $\Theta_X$ and $\Theta_Y$
\item $U_y$ is an $m_2 \times r$ matrix representing how the shared column structure is weighted over the rows of $\Theta_Y$
\item $V_z$ is an $n_2 \times r$ matrix representing how the shared row structure is weighted over the columns of $\Theta_Z$
%\item $S_x$, $S_y$, and $S_z$ are $r\times r$ scaling matrices for $\Theta_X$, $\Theta_Y$, and $\Theta_Z$, respectively.
%\item $E_x$, $E_y$ and $E_z$ are error matrices in which the entries are independent and have mean $0$.
\end{itemize}

To estimate the underlying decomposition, the GLMF algorithm uses a similar approach to the alternating least squares method that was used for LMF \citep{o2019linked}. It alternates between estimating the loadings $U$ and $U_y$ and the scores $V$ and $V_z$. However, the approach in that paper uses linear models for estimating the scores and loadings. As mentioned previously, this estimation approach is not as effective when the underlying data do not follow a normal distriibution. Instead, at each step the estimation of the scores or loadings is done using the IRLS algorithm described in \citep{li2018general}, which allows for heterogenous data types. This algorithm iteratively operates to maximize the likelihood of $\Theta_x$, $\Theta_Y$, and $\Theta_Z$ given $X$, $Y$, and $Z$, shown below.
\begin{align}
 L(\Theta \vert X,Y,Z) = \prod_{x_i \in X}{f_X(x_i \vert \theta_{X_i})} \prod_{y_i \in Y}{f_Y(y_i \vert \theta_{Y_i})} \prod_{z_i \in Z}{f_Z(z_i \vert \theta_{Z_i})}
\label{eq:glmflik}
\end{align}
In this likelihood, $f_X$, $f_Y$, and $f_Z$ are the probability density functions for the exponential family distributions assumed for $X$, $Y$, and $Z$. Expressing this equation in the general form for exponential family distributions gives the following lkelihood for the X matrix:
$$ L(\Theta_X \vert X) = \prod_{x_i \in X}{h_X(x_i)\exp{\{x_i\theta_{X_i} - b(\theta_{X_i})\}}}. $$ 
The likelihoods for $Y$ and $Z$ can be written similarly.

\subsection{IRLS Algorithm} \label{sec:irls}

The following steps describe the IRLS procedure for estimating the scores $V$ for a matrix $X$ given the loadings $U$, based on the heterogenous IRLS algorithm from \cite{li2018general}. Let $g()$ be the link function.
\begin{enumerate}
\item By default, all starting weights $w$ are set to 1, unless specified otherwise. 
\item Set the starting values for $\mu = X$. A small correction is made to binomial data so that the link function does not involve taking the log of zero, which is undefined. 
\item Set the natural parameter matrix $\Theta = g(\mu)$. For heterogeneous data, the link function may vary across rows or columns. In that case, the split the data into $n$ partitions, $i=1,..,n$, then set each $\Theta_i = g_i(\mu_i)$, where $g_i()$ is the link function corresponding to the distribution of the $i^{th}$ partition. 
\item Generate an induced response matrix $S = \Theta + \frac{X - \mu}{\frac{d\mu}{d\Theta}\Theta}$.
\item Set the weights $\tilde{w} = \sqrt{\frac{w[\frac{d\mu}{d\Theta}\Theta]^2}{Var(\mu)}}$. \\
Note: For heterogeneous data, steps 4 and 5 are partitioned by distribution (as in step 3).
\item For each row $i$, compute the weighted least squared estimate for $V_{i\cdot}$:
 $$ \hat{V_{i\cdot}} = (U^T \tilde{w}^2U)^{-1}U^T \tilde{w}^2S_{\cdot i} $$
\item Update $\Theta = X\hat{V}^T$.
\item Repeat Steps 4 through 7 until convergence
\end{enumerate}

A similar procedure is used for estimating $V$ given $U$. For simplicity in this case, $X^T$ is used as the response instead of $X$ to avoid transpositions within the algorithm, and the least squares estimate for each row $i$ of $\hat{U}$ is $ \hat{U_{i\cdot}} = (V^T \tilde{w}^2V)^{-1}V^T \tilde{w}^2S^*_{\cdot i} $, where $S^*$ is the induced response matrix based on $X^T$.

\subsection{GLMF Algorithm}

Given initial values, the algorithm proceeds by iteratively updating the components $U$, $V$, $U_y$, and $V_y$. First initialize $\tilde{V} = [V^T V_Z^T]^T$ as the first $r$ right singular vectors of the singular value decomposition (SVD) of $\tilde{Z}$.  The initial estimate for $V$ is the first $n_1$ rows of $\tilde{V}$.  Then repeat the following steps to maximize the likelihood in Equation \ref{eq:glmflik}: 
\begin{enumerate}
\item Update $U_y$ using the IRLS algorithm detailed above, given $Y$ and scores $V$.   % $ U_y = (V^T V)^{-1} V^TY $
\item Update $U$ via IRLS, given $\tilde{Z}$ and scores $\tilde{V}$.  % $ U = (\tilde{V}^T \tilde{V})^{-1} \tilde{V}^T \tilde{Z} $
%\item Scale $U$ by dividing each column by its Frobenius norm
\item Update $\tilde{U}$: $ \tilde{U} = [U^T \,\,\, U_y^T]^T $
\item Update $V$ via IRLS, given $\tilde{Y}$ and loadings $\tilde{U}$.     % $ V = (\tilde{U}^T \tilde{U})^{-1} \tilde{U}^T \tilde{Y} $
\item Update $V_z$ via IRLS, given $Z$ and loadings $U$.     % $ V_z = (U^T U)^{-1} U^T Z $
%\item Scale V by dividing each column by its Frobenius norm
%\item Update $S_x$ via least squares; define $W: np \times r$ such that the $i$'th column of $W$ is the vectorization of the product of the $i$th columns of $U$ and $V$, $W[,i] = \mbox{vec}(U[,i] V[,i]^T)$, then the diagonal entries of $S_x$ are $(W^T W)^{-1}W^T \mbox{vec}(X)$
\item Update $\tilde{V} = [V^T \,\,\, V_z^T]^T $.
\item Iterate Steps 1 through 6 until the estimates of $\mu$ converge. 
\end{enumerate}
The variance estimate $\hat{\sigma}^2$ is also updated to maximize the likelihood for the normally distributed data at each stage. The algorithm results in the following rank $r$ estimates for the joint structure natural parameters:
\begin{align}
\label{facEq}
\begin{split}
 \Theta_X &= UV^T \\
 \Theta_Y &= U_yV^T  \\
 \Theta_Z &= UV_z^T.
 \end{split}
\end{align}

\subsection{Illustrative Simulation} \label{ill_sim}

To test the GLMF algorithm's ability to recover the underlying natural parameters, a single simulation was run with three data sources $X$, $Y$, and $Z$. Rank 3 scores and loadings were simulated from a N(0,0.4) distribution. These scores and loadings were used to construct the natural parameters $\Theta$ for the simulated $X$, $Y$, and $Z$ matrices. A binomial sample size matrix $N$ was generated from integer values between 1 and 8, inclusive. An inverse logit transformation was used to calculate true binomial probabilities $p$ from $\Theta_X$, and $X$ was drawn from a Bin(N, p) distribution. Y and Z were drawn from N($\Theta_Y$,0.1) and N($\Theta_Z$,0.1), respectively. A rank 3 GLMF model was fit to this simulated dataset. Figure 2 shows that this low-rank approximation of the data does a fairly good job of recovering the true values for $p$, $\mu_y = \Theta_y$, and $\mu_z = \Theta_z$, with the estimated values falling close to the diagonal. The Pearson correlation values between the true and estimated values of $p$, $\mu_y$, and $\mu_z$ were 0.979, 0.993, and 0.992, respectively.

\section{Imputation} 
\label{sec:impute}

The overall goal of this method in this paper is to be able to predict batting averages of specific matchups. To generate these predictions, I used an imputation method that alternates between estimating the GLMF components based on the data and predicting the batting averages using those components. 

\cite{o2019linked} described an algorithm for imputing data using the LMF algorithm. In this paper, I used a similar algorithm, imputing values for the $X$ matrix by alternating between imputing $X$ using the GLMF scores and loadings and estimating the GLMF decomposition using the imputed values. In this case, the $X$ matrix contains binomial data instead of normally distributed data, so instead of generating imputed missing values, I used the imputed binomial probabilities.

\subsection{Algorithms}
\label{imputeAlg}

Six different imputation methods were compared for predicting batting averages. The first method is the most naive approach, in which the prediction for each batter is just the overall mean batting average. The second method is based on the log5 statistic \citep{james1983abs}. Defining $B$ as the overall batting average for a batter, $P$ as the overall batting average for a pitcher, and $T$ as the overall batting average across all batters and pitchers, then the log5 estimate is defined as $\hat{p} = \frac{P \times B}{T} $.

The four remaining methods were all based on dimension reduction methods. Two of the methods (LMF and GLMF) incorporate the additional data sources of pitching and batting data. The other two methods only use information from the batter vs. pitcher matrix ($X$). One of these is principal components analysis (PCA), which is fit using a singular value decomposition of the centered and scaled matrix. PCA, like LMF, assumes normally distributed data, so I also considered an exponential PCA \citep{collins2002generalization}. I used the logisticPCA package to fit a logistic PCA model (LPCA), which is an implementation of exponential PCA for binomial data \citep{landgraf2015}.

The imputation algorithms for the dimension reduction methods are based on the iterative algorithm described in \cite{o2019linked}. It is worth noting that for matchups with no data, I treated the number of at-bats ($N$) as 1. The dimension reduction imputation methods all operate similarly as follows:
\begin{enumerate}
\item Begin by initializing the missing values. To initialize $\hat{p}_{ij}$, the estimated probability matrix for $X$, average the observed probabilities for each row ($\hat{p}_{i \cdot}$) and each column ($\hat{p}_{\cdot j}$). To compute each $\hat{p}_{ij}$, take the average of $\hat{p}_{i \cdot}$ and $\hat{p}_{\cdot j}$. Replace the missing values of $X$ with $\hat{p}$, and use N=1 for all of the missing values. 
\item Fit one of the dimension reduction methods (GLMF, LMF, LPCA, or PCA) using the imputed $\hat{X}$ matrix. For LMF and PCA, use the imputed probability matrix $\hat{p}$ instead of $\hat{X}$. 
\item Create $\hat{X}$ by replacing the missing values in $X$ with the fitted values from the dimension reduction. 
\item Repeat steps 2 and 3 until the successive estimates of $\hat{X}$ converge. 
\end{enumerate}

\subsection{Simulation}
\label{sec:glmf_imp_sim}

To compare the imputation algorithms, I simulated 144 data sets with 3 different varying parameters. The first parameter, $\sigma$, controlled the standard deviation of the model components. The values of this parameter were 0.1, 0.3, 0.5, and 0.7. This represents the amount of systematic variation. In terms of batting averages, higher values of sigma correspond to more variability in the true batting averages. This is not to be confused with the error standard deviation, which was fixed at 0.3. The next parameter, $nmax$, controlled the maximum number of at-bats for each matchup. The values chosen for this parameter were 1, 2, 8, and 16. The third parameter, $r$, controlled the rank of the components used to generate the data. The values chosen for rank were 1, 2, and 3. There are 48 combinations of these parameters, and each combination was repeated three times, leading to the 144 total simulated data sets. For each simulation, the dimension of $X$ was 200 $\times$ 200, the dimension of $Y$ was 50 $\times$ 200, and the dimension of $Z$ was 200 $\times$ 50. 

For each simulation, scores and loadings were drawn from a Normal(0, $\sigma ^2$) distribution. Then the natural parameter matrices ($\Theta_X$, $\Theta_Y$, $\Theta_Z$) were generated from the model in Equation \ref{facEq}. Binomial probabilities $p$ were generated by applying the inverse logit transformation to $\Theta_X$. The number of trials $N$ were selected at random from 1 to $nmax$. The entries for $X$ were generated from a Binomial($N$, $p$) distribution. The entries for $Y$ and $Z$ were simulated from Normal($\Theta_Y$, 0.09) and Normal($\Theta_Z$, 0.09) distributions, respectively. Then 20\% of the observations in each data set were set to missing, and the six imputation methods were applied to each to estimate the batting averages $p$.
 
Two different metrics were used to compare the performance of the methods. One method was root mean squared error (RMSE). RMSE is a common metric for evaluating cross-validation error, but it tends to favor the methods that are based on a normal distribution, since squared error loss is the loss function for those models. So the log likelihood under a binomial distribution was also computed for the predicted true batting averages given the observed batting averages. The GLMF algorithm failed to converge for two of the simulated data sets. These were both at rank 3 with lower variability, so these were likely the result of singular matrices in the underlying regression models. 

Figure 3 shows the average simulation results for each of the models against each of the three parameters that varied. Each of these is averaged over the other two parameters. Full results from this simulation are shown in Table \ref{tab:rmse16} and Table \ref{tab:logl16}. These are shown at a fixed value of $nmax$ because the maximum number of at-bats ($nmax$) did not appear to affect the relative performances of the imputation algorithms (Figure 3a). When the systematic variability is low ($\sigma=0.1$), then the mean imputation method works best because there is very little variation in the resulting batting averages $p$. However, the other methods perform better as the systematic variability increased. At $\sigma=0.5$ and $\sigma=0.7$, GLMF was the best-performing method with respect to log-likelihood. At $\sigma=0.7$, GLMF even outperformed the other methods in terms of RMSE, a metric that should favor the Gaussian methods. The LMF method also worked well and was generally the best method at $\sigma=0.3$. 

 \begin{table}[!ht]
  %\centering
    \caption[Mean RMSE for different imputation methods]{Mean RMSE for each of the six imputation methods on data simulated with rank $r$ scores and loadings, each generated from a normal distribution with standard deviation $\sigma$. The results shown use $nmax=16$. The GLMF algorithm failed to converge in two of the simulations; these entries are marked with asterisks and represent the means of two replicates instead of three.}
  \begin{tabular}{ c c | c c c c c c }
    %\hline
     $r$ & $\sigma$ & GLMF & LMF & LPCA & PCA & Log5 & Mean \\
      \hline
      1 &  0.1 &  0.0240 & 0.0246 & 0.0205 & 0.0350 & 0.0225 & 0.0019 \\
      1 &  0.3 &  0.0285 & 0.0114 & 0.0262 & 0.0360 & 0.0277 & 0.0160  \\
      1 &  0.5 &  0.0200 & 0.0122 & 0.0495 & 0.0302 & 0.0504 & 0.0453  \\
      1 &  0.7 &  0.0188 & 0.0202 & 0.0913 & 0.0321 & 0.0916 & 0.0899  \\
      2 &  0.1 &  0.0339 & 0.0340 & 0.0289 & 0.0464 & 0.0223 & 0.0029 \\
      2 &  0.3 &  0.0314 & 0.0164 & 0.0358 & 0.0474 & 0.0324 & 0.0243  \\
      2 &  0.5 &  0.0235 & 0.0191 & 0.0557 & 0.0393 & 0.0703 & 0.0668  \\
      2 &  0.7 &  0.0223 & 0.0302 & 0.0951 & 0.0421 & 0.1280 & 0.1270   \\
      3 &  0.1 &  0.0416* & 0.0438 & 0.0352 & 0.0571 & 0.0227 & 0.0037 \\
      3 &  0.3 &  0.0359* & 0.0211 & 0.0440 & 0.0576 & 0.0365 & 0.0289  \\
      3 &  0.5 &  0.0270 & 0.0239 & 0.0598 & 0.0460 & 0.0828 & 0.0796  \\
      3 &  0.7 &  0.0259 & 0.0380 & 0.0967 & 0.0514 & 0.1530 & 0.1520   \\
    %\hline
  \end{tabular}
  \label{tab:rmse16}
\end{table}

 \begin{table}[!ht]
  %\centering
    \caption[Mean log likelihood for different imputation methods]{Mean log likelihood for each of the six imputation methods on data simulated with rank $r$ scores and loadings, each generated from a normal distribution with standard deviation $\sigma$. The results shown use $nmax=16$. The GLMF algorithm failed to converge in two of the simulations; these entries are marked with asterisks and represent the means of two replicates instead of three.}
  \begin{tabular}{ c c | c c c c c c }
    %\hline
     $r$ & $\sigma$ & GLMF & LMF & LPCA & PCA & Log5 & Mean \\
      \hline
      1 &  0.1 & -12178 & -12194 & -12153 & -12322 & -12170 & -12081 \\
      1 &  0.3 & -12225 & -12101 & -12189 & -12328 & -12205 & -12125 \\
      1 &  0.5 & -12143 & -12126 & -12556 & -12269 & -12561 & -12480 \\
      1 &  0.7 & -11982 & -12243 & -13349 & -12311 & -13358 & -13315 \\
      2 &  0.1 & -12307 & -12308 & -12242 & -12497 & -12164 & -12074 \\
      2 &  0.3 & -12272 & -12151 & -12314 & -12519 & -12262 & -12191 \\
      2 &  0.5 & -12053 & -12095 & -12519 & -12345 & -12839 & -12754  \\
      2 &  0.7 & -11827 & -12395 & -13276 & -12509 & -14457 & -14454  \\
      3 &  0.1 & -12467*& -12488 & -12346 & -12830 & -12197 & -12103 \\
      3 &  0.3 & -12276*& -12142 & -12427 & -12746 & -12302 & -12209 \\
      3 &  0.5 & -12082 & -12158 & -12575 & -12491 & -13131 & -13026 \\
      3 &  0.7 & -11631 & -12403 & -13079 & -12519 & -15475 & -15454 \\
    %\hline
  \end{tabular}
  \label{tab:logl16}
\end{table}

\section{Cross-validation Study}

Aggregate batting data, aggregate pitching data, and batter vs. pitcher data for the 2017 season were obtained from MLB.com. The data was filtered to include only pitchers with more than 20 innings pitched and batters with at least 50 at-bats in the season. The resulting data set contained 516 pitchers and 508 batters. The aggregate pitching data included the following statistics: wins, losses, games, games started, games finished, complete games, shutouts, saves, innings pitched, hits, runs, earned runs, homeruns, walks, intentional walks, strikeouts, hit by pitch, balks, and wild pitches. These statistics were scaled by each pitcher's batters faced, so they represent a proportion. The aggregate batting data included the following batting statistics: games, at-bats, runs, hits, doubles, triples, homeruns, runs batted in, stolen bases, caught stealing, walks, strikeouts, total bases, grounded into double play, hit by pitch, sacrifice hits, sacrifice flies, and intentional walks. These values were similarly scaled by the number of plate appearances for each batter. 

A five-fold cross-validation procedure was used to evaluate the ability of the different dimension reduction methods to predict batting averages in specific batter/pitcher matchups. Of the 262,128 possible matchups (516 pitchers times 508 batters), at least one at-bat was observed for 62,528 of the matchups. This means that empirical batting averages were observed for about 24 percent of the possible matchups. Those 62,528 matchups were split into five testing sets, each containing observed batting averages for about 12,506 matchups. In each cross-validation fold, the testing observations were set to missing. Then, the batting averages were imputed based on the remaining observations using each of the five imputation methods described in Section \ref{sec:impute}. This procedure was repeated using different ranks for each of the dimension reduction methods. Ranks 1, 2, and 3 were used for each of the methods. 

Two metrics were used to assess the accuracy of the predictions. One metric was the mean squared error (MSE) of the predictions. However, the binomial models are not based on a squared error loss function. So the other metric that was used was the log likelihood of observing $X$ hits in $N$ at-bats for each matchup given the imputed batting averages $\hat{p}$. Lower and upper bounds of 0.001 and 0.999 were placed on the imputation methods, since the LMF and SVD methods are not contrained to the space of (0,1). Otherwise, estimated probabilities less than zero or greater than one would result in log likelihoods that are undefined. 

The MSE values and the log likelihoods of each of the imputation procedures are summarized in Table \ref{tab:cv_results}. GLMF consistently performed better than all the other methods, as it had both the lowest MSE and highest log likelihood values based on the cross-validation procedure. However, the naive mean imputation approach was the next best method in terms of performance. This could be partially because the mean imputation value of 0.250 is likely a reasonable estimate for most batting averages because batting averages in the long run tend to fall in the range of 0.200 to 0.300, which is a fairly narrow range. An uninformative method outperforming the other methods indicates that they may be overfitting the observed data, resulting in high bias. This is supported by the fact that an increase in model complexity (by increasing the rank of the approximation) actually worsened the performance of the LMF, LPCA, and PCA models. In contrast, increasing the complexity of the GLMF method actually improved the performance, with a slight increase in the log-likelihood. 

A graphic representation of the estimated batting averages is given in Figure 4. This figure shows smoothed LOESS curves representing the relationship between the observed batting averages and each of the imputation methods. One interesting feature of the estimated batting average curves for all of the methods is that they are negatively associated with the observed batting averages above observed averages of about 0.667. However, there is a reasonable explanation for this. Many of the high observed batting averages are the product of a single hit in a single at-bat. One observed data point cannot give an accurate estimate of a batting average for a certain matchup, since no matchup is guaranteed to always result in a hit or an out. So in these cases, the model predictions are probably more reasonable estimates of batting averages in future matchups than the current empirical estimates (which can only be 0.000 or 1.000 with only a single observation). Another interesting note is that most of the methods seem to systematically overestimate the averages up until about 0.265, but the LPCA method seems to systematically underestimate the averages instead. It is possible that the performance of these methods could be improved by applying some form of bias correction to account for these issues. 

Based on the rank 3 GLMF results, the matchup that was most favorable in 2017 was Jose Altuve vs. Bartolo Colon, with a predicted average of 0.463. They actually faced each other three times in 2017, and Altuve had one hit, so the observed batting average was 0.333. It is not surprising that Altuve would be involved in the most favorable matchup because he was the American League most valuable player in 2017. 

 \begin{table}
    \caption[Cross-validation results of the imputed values of the true batting average $p$]{Cross-validation results for predicting batting averages using each of the six imputation methods. The dimension reduction methods were tested at rank 1, 2, and 3. Results are presented for both RMSE and log likelihood.} 
  \label{tab:cv_results}
  \begin{tabular}{ c c c c c c c }
    RMSE & GLMF & LMF & LPCA & PCA & Log5 & Mean  \\
    \hline
    Rank 1 & 0.342 & 0.344 & 0.343 & 0.348 & 0.345 & 0.344 \\
    Rank 2 & 0.342 & 0.345 & 0.344 & 0.353 & - & -  \\
    Rank 3 & 0.342 & 0.351 & 0.344 & 0.358 & - & - \\
    \hline
    log L & GLMF & LMF & LPCA & PCA & Log5 & Mean  \\
    \hline
    Rank 1 & -0.857 & -0.865 & -0.862 & -0.885 & -0.870 & -0.864 \\
    Rank 2 & -0.856 & -0.866 & -0.863 & -0.916 & - & - \\
    Rank 3 & -0.854 & -0.899 & -0.864 & -0.952 & - & - \\
  \end{tabular}
\end{table}

\section{Discussion}

As illustrated through simulations and cross-validation with real MLB data, the GLMF algorithm can be used to impute missing batting averages while incorporating other sources of data. As the model complexity increased, the GLMF method was the only method that improved in terms of imputation accuracy. The GLMF method at rank 3 had the most accurate imputed batting averages in the cross-validation study. The gain in imputation accuracy relative to LPCA was small, but the real advantage of the GLMF method is its flexibility. It is important to remember that this analysis used relatively uninformative data for $Y$ and $Z$. While the aggregate batting and pitching stats can give some insight into tendencies of certain batters and pitchers, advances in technology have made it possible to collect much more informative data, such as the Statcast data collected by the MLB. This data could be incorporated into the GLMF method to improve prediction accuracy. If more informative data is used for $Y$ and $Z$, then the performance of GLMF will be even further ahead of the other imputation methods. This can be seen in the simulation study: as $\sigma$ increases, the GLMF method performs better relative to the other methods. 

This paper focused on a specific context, in which $X$ followed a binomial distribution, and $Y$ and $Z$ followed a normal distribution. But the GLMF algorithm is applicable to any exponential family distributions in any combination. Although it is primarily useful for heterogeneously distributed data, it can also be used for homogeneously distributed data, such as all normally distributed or all binomially distributed data sets. In the case of normally distributed data sets, this simplifies to an LMF model, which is a special case of GLMF. Although this analysis only used three matrices, with the entries of each following a single distribution, these methods are not limited to this strict data structure. The heterogeneous IRLS algorithm used to compute joint structure allows for any number of different distributions. This allows for both modeling more than two horizontally or vertically integrated data sets simultaneously even if they all follow different distributions. For instance, the batter vs. pitcher data, aggregate batting data, and Statcast batting data could all be incorporated into the GLMF model with common loadings but unique scores. It also allows the entries within a matrix to follow different distributions when one of the matrices has variables that follow different distributions.

%\begin{acknowledgement}
%  Please insert acknowledgments of the assistance of colleagues or similar notes of appreciation here.
%\end{acknowledgement}

\begin{funding}
  None declared.
\end{funding}

\bibliographystyle{chicago}
\bibliography{glmf}

\begin{thebibliography}{}

\bibitem[\protect\citeauthoryear{Collins, Dasgupta, and Schapire}{Collins
  et~al.}{2002}]{collins2002generalization}
Collins, M., S.~Dasgupta, and R.~E. Schapire (2002).
\newblock A generalization of principal components analysis to the exponential
  family.
\newblock In {\em Advances in neural information processing systems}, pp.\
  617--624.

\bibitem[\protect\citeauthoryear{Doo and Kim}{Doo and
  Kim}{2018}]{doo2018modeling}
Doo, W. and H.~Kim (2018).
\newblock Modeling the probability of a batter/pitcher matchup event: A
  bayesian approach.
\newblock {\em Plos one\/}~{\em 13\/}(10), e0204874.

\bibitem[\protect\citeauthoryear{Green}{Green}{1984}]{green1984iteratively}
Green, P.~J. (1984).
\newblock Iteratively reweighted least squares for maximum likelihood
  estimation, and some robust and resistant alternatives.
\newblock {\em Journal of the Royal Statistical Society. Series B
  (Methodological)\/}, 149--192.

\bibitem[\protect\citeauthoryear{Healey}{Healey}{2015}]{healey2015modeling}
Healey, G. (2015).
\newblock Modeling the probability of a strikeout for a batter/pitcher matchup.
\newblock {\em IEEE Transactions on Knowledge and Data Engineering\/}~{\em
  27\/}(9), 2415--2423.

\bibitem[\protect\citeauthoryear{James}{James}{1983}]{james1983abs}
James, B. (1983).
\newblock {\em The Bill James Baseball Abstract}.
\newblock New York, NY: Ballantine Books.

\bibitem[\protect\citeauthoryear{Landgraf and Lee}{Landgraf and
  Lee}{2015}]{landgraf2015}
Landgraf, A.~J. and Y.~Lee (2015).
\newblock Dimensionality reduction for binary data through the projection of
  natural parameters.
\newblock Technical Report 890, Department of Statistics, The Ohio State
  University.

\bibitem[\protect\citeauthoryear{Lewis}{Lewis}{2004}]{lewis2004moneyball}
Lewis, M. (2004).
\newblock {\em Moneyball: The art of winning an unfair game}.
\newblock WW Norton \& Company.

\bibitem[\protect\citeauthoryear{Li and Gaynanova}{Li and
  Gaynanova}{2018}]{li2018general}
Li, G. and I.~Gaynanova (2018).
\newblock A general framework for association analysis of heterogeneous data.
\newblock {\em The Annals of Applied Statistics\/}~{\em 12\/}(3), 1700--1726.

\bibitem[\protect\citeauthoryear{Lock, Hoadley, Marron, and Nobel}{Lock
  et~al.}{2013}]{lock2013joint}
Lock, E.~F., K.~A. Hoadley, J.~S. Marron, and A.~B. Nobel (2013).
\newblock Joint and individual variation explained ({JIVE}) for integrated
  analysis of multiple data types.
\newblock {\em The annals of applied statistics\/}~{\em 7\/}(1), 523.

\bibitem[\protect\citeauthoryear{L{\"o}fstedt and Trygg}{L{\"o}fstedt and
  Trygg}{2011}]{lofstedt2011onpls}
L{\"o}fstedt, T. and J.~Trygg (2011).
\newblock Onpls—a novel multiblock method for the modelling of predictive and
  orthogonal variation.
\newblock {\em Journal of Chemometrics\/}~{\em 25\/}(8), 441--455.

\bibitem[\protect\citeauthoryear{O'Connell and Lock}{O'Connell and
  Lock}{2019}]{o2019linked}
O'Connell, M.~J. and E.~F. Lock (2019).
\newblock Linked matrix factorization.
\newblock {\em Biometrics\/}~{\em 75\/}(2), 582--592.

\bibitem[\protect\citeauthoryear{Park and Lock}{Park and
  Lock}{2020}]{park2020integrative}
Park, J.~Y. and E.~F. Lock (2020).
\newblock Integrative factorization of bidimensionally linked matrices.
\newblock {\em Biometrics\/}~{\em 76\/}(1), 61--74.

\bibitem[\protect\citeauthoryear{Yang and Michailidis}{Yang and
  Michailidis}{2016}]{yang2016non}
Yang, Z. and G.~Michailidis (2016).
\newblock A non-negative matrix factorization method for detecting modules in
  heterogeneous omics multi-modal data.
\newblock {\em Bioinformatics\/}~{\em 32\/}(1), 1--8.

\bibitem[\protect\citeauthoryear{Yuan and Gaynanova}{Yuan and
  Gaynanova}{2021}]{yuan2021double}
Yuan, D. and I.~Gaynanova (2021).
\newblock Double-matched matrix decomposition for multi-view data.
\newblock {\em arXiv preprint arXiv:2105.03396\/}.

\bibitem[\protect\citeauthoryear{Zhou, Cichocki, Zhang, and Mandic}{Zhou
  et~al.}{2015}]{zhou2015group}
Zhou, G., A.~Cichocki, Y.~Zhang, and D.~Mandic (2015).
\newblock Group component analysis for multiblock data: Common and individual
  feature extraction.
\newblock {\em IEEE transactions on neural networks and learning systems\/}.

\end{thebibliography}

\newpage
\includegraphics[scale=0.5]{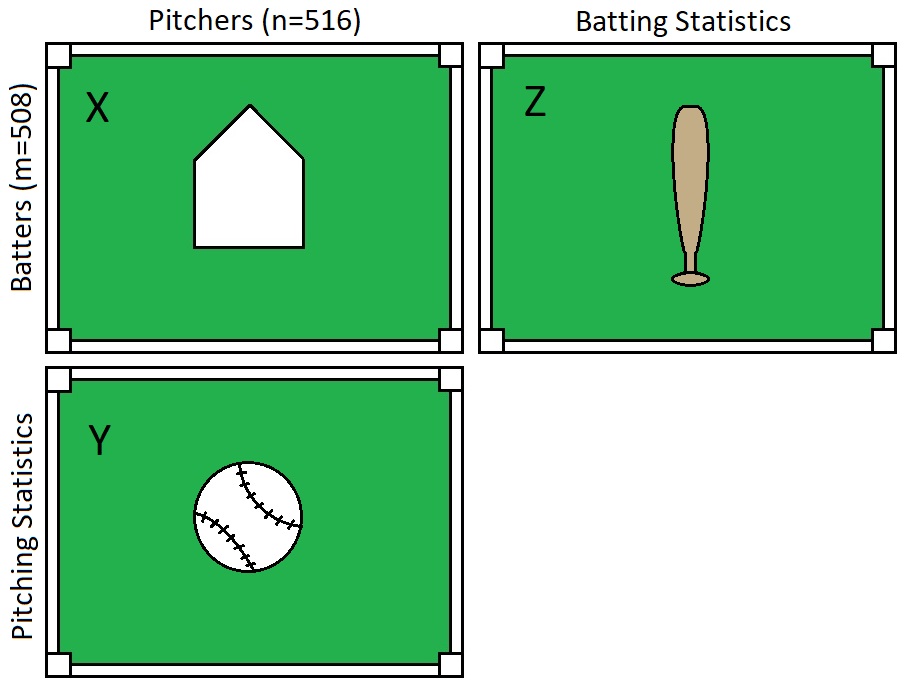}

\small{\textbf{Figure 1.} An illustration of the structure of the data. The batter vs. pitcher data is the $X$ matrix, indicated with a homeplate. This matrix contains counts of the number of hits for each matchup. The $N$ matrix, which is not shown, has the same dimensions as $X$ and contains the number of at-bats for each matchup. The pitching data $Y$, indicated with a baseball, shares its sample set (pitchers) with the batter vs. pitcher data data. Likewise, the batting data $Z$, indicated with a bat, shares its sample set (batters) with the batter vs. pitcher data.} 

\newpage
\includegraphics[scale=0.8]{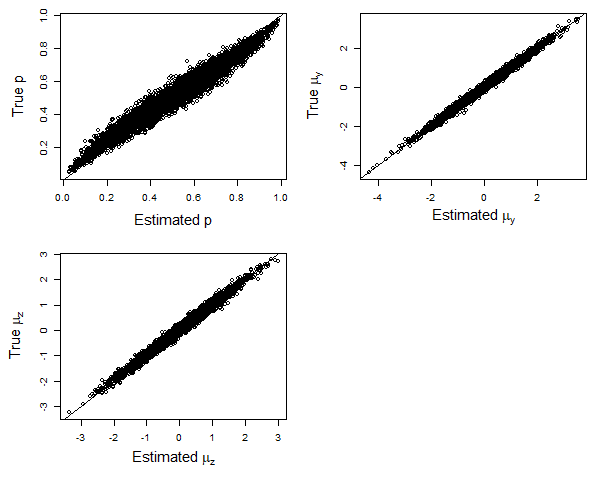}

\small{\textbf{Figure 2.} Estimated mean parameters of the GLMF decomposition for a single simulated data set compared to the true parameter values. For X, this is the estimated binomial probability for each entry in the matrix. For Y and Z, this is the estimated normal mean for each entry.}

\newpage
\includegraphics[scale=0.35]{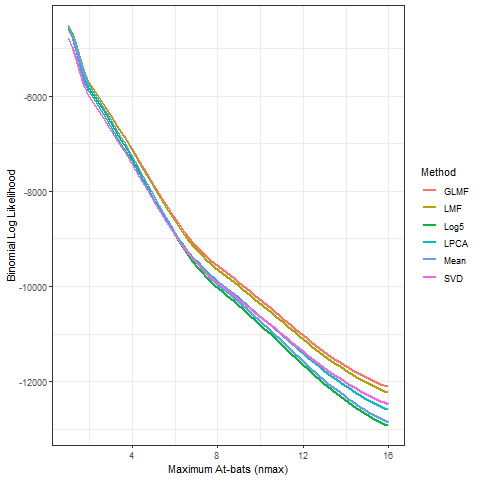}
\includegraphics[scale=0.35]{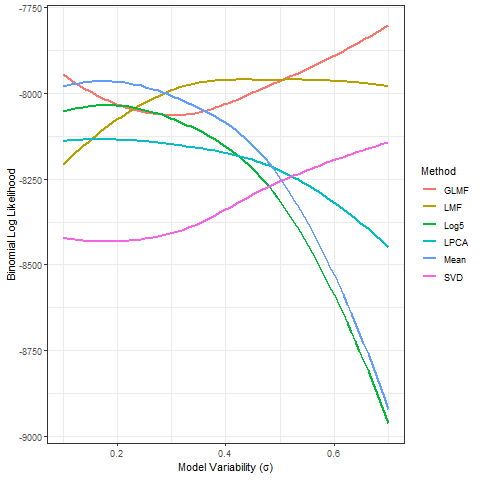}
\includegraphics[scale=0.35]{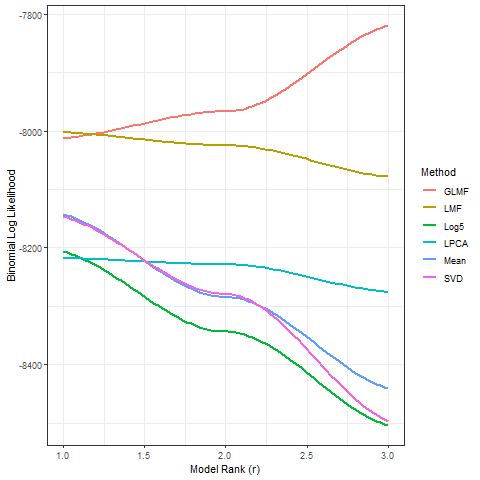}

\small{\textbf{Figure 3.} Relative performance of the six imputation methods from the simulation study. \textbf{(a)} Binomial log-likelihood as a function of the maximum number of at-bats (nmax), averaged over variability and rank. \textbf{(b)} Binomial log-likelihood as a function of the systematic variability in the model, averaged over nmax and rank. \textbf{(c)} Binomial log-likelihood as a function of the rank of the underlying model, averaged over nmax and variability.}

\newpage
\includegraphics[scale=0.4]{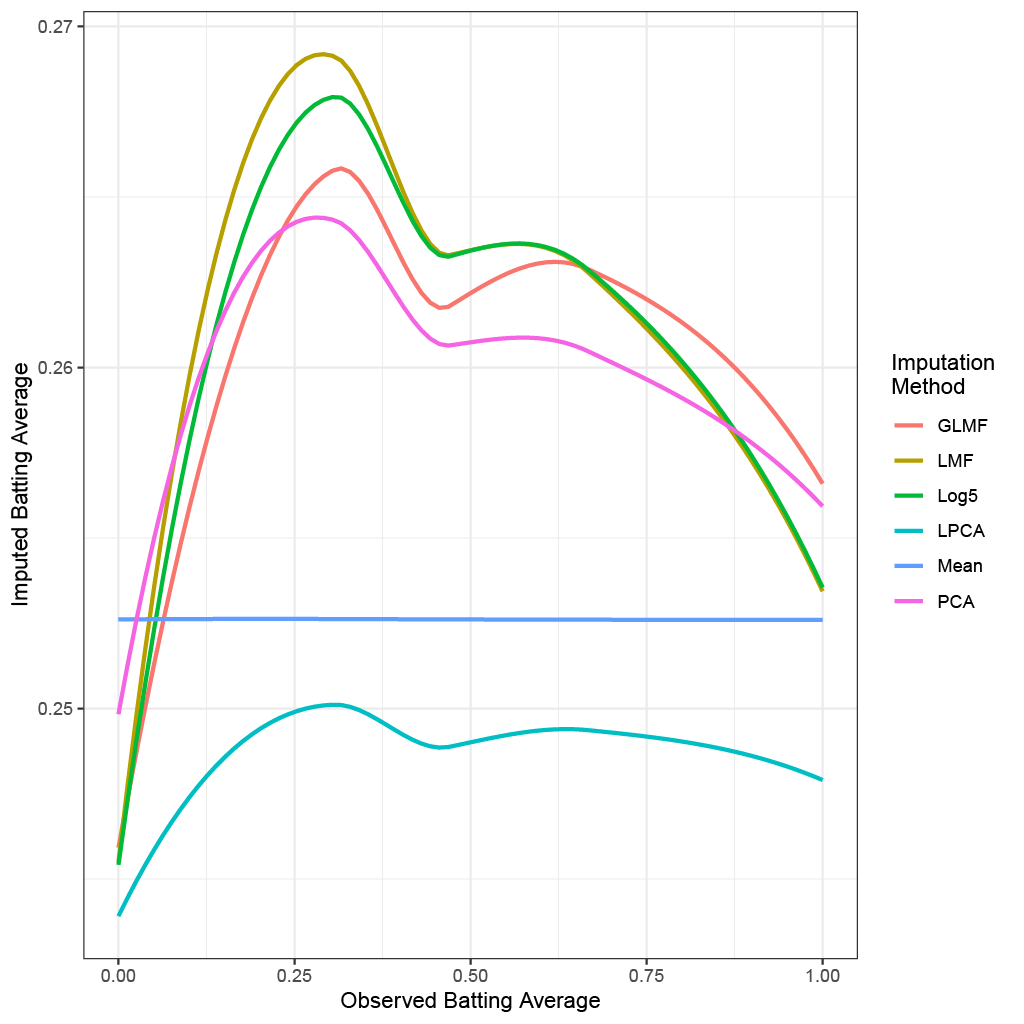}

\small{\textbf{Figure 4.} Predicted batting averages from each of the imputation methods compared to the observed batting averages, based on the five-fold cross-validation study. For each of the dimension reduction methods, only the results from the rank 3 approximation are shown.}

\end{document}